\documentclass[10pt,aps,prd,onecolumn,showpacs,amsmath,amssymb,nofootinbib,eqsecnum,preprintnumbers,longbibliography,superscriptaddress]{revtex4-2}

\usepackage[english]{babel}				
\usepackage{amsmath}
\usepackage{amsthm}
\usepackage{mathtools}

\usepackage{xcolor}

\usepackage{tcolorbox}
\usepackage{pagecolor}

\usepackage[colorlinks=true,
            citecolor=red,
            linkcolor=blue,
            urlcolor=violet,
            filecolor=cyan,
            backref=false]{hyperref}

\newtheorem*{thm}{Theorem}

\newcommand\bs[1]{\boldsymbol{#1}}

\newcommand\dd{\mathrm{d}}
\newcommand\pp{\partial}

\DeclareMathOperator\sech{sech}
\DeclareMathOperator\arcsinh{arcsinh}

\newcommand\feq{\mathrel{\phantom{=}}}

\newtcolorbox{ivanbox}{fontupper=\scshape}

\DeclareMathOperator*{\sumint}{%
\mathchoice%
  {\ooalign{$\displaystyle\sum$\cr\hidewidth$\displaystyle\int$\hidewidth\cr}}
  {\ooalign{\raisebox{.14\height}{\scalebox{.7}{$\textstyle\sum$}}\cr\hidewidth$\textstyle\int$\hidewidth\cr}}
  {\ooalign{\raisebox{.2\height}{\scalebox{.6}{$\scriptstyle\sum$}}\cr$\scriptstyle\int$\cr}}
  {\ooalign{\raisebox{.2\height}{\scalebox{.6}{$\scriptstyle\sum$}}\cr$\scriptstyle\int$\cr}}
}

\begin{document}

\title{New non-singular cosmological solution of non-local gravity}

\author{Ivan Kol\'a\v{r}}
\email{i.kolar@rug.nl}
\affiliation{Van Swinderen Institute, University of Groningen, 9747 AG, Groningen, Netherlands}
\author{Francisco Jos\'e Maldonado Torralba}  \email{fmaldo01@ucm.es}
\affiliation{Cosmology and Gravity Group, Department of Mathematics and Applied Mathematics, University of Cape Town, Rondebosch 7701, Cape Town, South Africa}
\author{Anupam Mazumdar}
\email{anupam.mazumdar@rug.nl}
\affiliation{Van Swinderen Institute, University of Groningen, 9747 AG, Groningen, Netherlands}

\begin{abstract}
We present a new bouncing cosmological solution of the non-local theory known as infinite derivative gravity, which goes beyond the recursive ansatz, ${\Box R = r_1 R +r_2}$. The non-local field equations are evaluated using the spectral decomposition with respect to the eigenfunctions of the wave operator. The energy-momentum tensor computed for this geometry turns out to be much more sensitive to the choice of the non-local form-factor, since it depends on the value of the function on a continuous infinite interval. We show that this stronger dependence on the form-factor allows us to source the geometry by the perfect fluid with the non-negative energy density satisfying the strong energy condition. We show that this bouncing behaviour is not possible in the local theories of gravity such as in general relativity or $R+R^2$ gravity sourced by a fluid which meets the non-negative energy and strong energy conditions.
\end{abstract}

\maketitle

\section{Introduction}

There is no doubt that the current cosmological model, $\Lambda$ cold dark matter ($\Lambda$CDM), gives so far the most accurate description of the history of the universe \cite{Planck:2015fie}. Nevertheless, it suffers from some important shortcomings, such as the tensions between early and late time measures \cite{Riess:2019qba}, the nature of dark energy \cite{Weinberg:1988cp,Martin:2012bt}, and the initial singularity problem \cite{Novello:2008ra,Senovilla:2018aav}. The latter is the most worrisome from a conceptual perspective, since the existence of singularities signals the limited range of validity of the theory. 

By the time the Friedmann--Lema\^itre--Robertson--Walker (FLRW) cosmological solution was introduced, `${t=0}$ singularity' was thought to be a consequence of spacetime symmetries. It was not until the development of the singularity theorems in 1960s by Penrose and Hawking \cite{Penrose:1964wq,Hawking:1967ju,Hawking:1969sw,Hawking:1973uf} that the sufficient conditions for the appearance of singularities were elucidated. In particular, in a cosmological context one can recall the Hawking singularity theorem, which states \cite{Hawking:1967ju}:
\begin{thm}
Let $M$ be a 4-dimensional manifold equipped with the metric $\bs{g}$ satisfying the following conditions:
\begin{enumerate}
    \item Global hyperbolicity.
    \item Temporal convergence condition, i.e. ${\bs{v}\cdot\bs{Ric}\cdot\bs{v}\ge 0}$ for every timelike vector $\bs{v}$, where $\bs{Ric}$ is the Ricci tensor and $\cdot$ denotes a contraction between adjacent tensor indices.
    \item There exists a Cauchy hypersurface $\Sigma$ with expansion of the future-directed geodesic congruence greater or equal than a constant $A>0$.
\end{enumerate}
Then no past-directed timelike curve from $\Sigma$ can have a length greater than ${3/\left|A\right|}$. In particular, all past-directed timelike geodesics are incomplete.
\end{thm}

It is clear that this theorem predicts the initial singularity for quite general requirements, and hence this behavior is not due to the specific symmetry of the considered solution. When applied to a FLRW metric that describes the expansion of the universe, the first and third condition hold. Therefore, in order to predict singularities in such spacetimes, we just need to worry about the second condition in the theorem, the \textit{temporal convergence condition}. Such a condition is completely theory-independent, but it can be related to the energy-momentum tensor of a specific theory via the field equations, obtaining what is known as the \textit{energy conditions}.

In the \textit{general relativity (GR)}, the temporal convergence condition written in terms of the energy-momentum tensor $\bs{T}$ can be expressed as 
\begin{equation}
    \bs{v}\cdot\left(\bs{T}-\tfrac12 T \bs{g}\right)\cdot \bs{v}\geq 0\;,
\end{equation}
for every timelike vector $\bs{v}$. Such a condition is usually referred to as the \textit{strong energy condition (SEC)}. In a particular case of the perfect fluid, which is a source of a FLRW spacetime, SEC is equivalent to
\begin{equation}
    \rho+p\ge0\;, \quad\rho+3p\ge0\;,
\end{equation}
where $\rho$ and $p$ are the \textit{energy density} and the \textit{pressure} of the perfect fluid, respectively. SEC has an important physical interpretation: it guarantees the attractive character of gravity in GR \cite{Hawking:1973uf,Wald:1984rg}.

Because of issues linked to the initial singularity problem, the non-singular cosmological solutions have become a widely studied topic in gravitation (for a comprehensive review, see \cite{Novello:2008ra}). The non-singular cosmological solutions which contract and then expand are called the \textit{bouncing cosmologies}. The first explicit solutions were obtained in late 1970s by Novello and Salim \cite{Novello:1979ik} and Melnikov and Orlov \cite{Melnikov:1979hf}. Nevertheless, at that time they did not attract too much attention of the community, since they require the violation of SEC. It was not until two decades later, with the measurement of the accelerated expansion of the universe, that these kinds of metrics started to be regarded as possible physical solutions of the problems present in the standard cosmological models.

Resorting to Hawking's theorem, it is clear that in order to obtain non-singular bouncing solutions in GR, SEC must be necessarily violated \cite{Molina-Paris:1998xmn,Cattoen:2005dx} implying that gravity is not attractive at the bounce. This fact has served as motivation to explore bouncing solutions in modified theories of gravity, where the unattractive character can be a consequence of the modification, not of the energy-momentum tensor. For a review of bouncing universes in various theories of gravity we refer the reader to~\cite{Nojiri:2017ncd}.

In this paper, we are interested in non-local theory of gravity known as the \textit{infinite derivative gravity (IDG)} \cite{Krasnikov:1987yj,kuzmin1989finite,Biswas:2005qr,Biswas:2011ar}, which contains operators with an infinite number of derivatives in the action. The most general action (quadratic in curvature) with no ghosts or extra degrees of freedom has been constructed around Minkowski~\cite{Biswas:2011ar}, de Sitter, and anti-de Sitter spacetimes~\cite{Biswas:2016etb}. The non-local gravitational interaction has been argued to improve ultraviolet aspects gravity at the quantum level~\cite{Tomboulis:1997gg,Modesto:2011kw,Talaganis:2014ida,Abel:2019ufz,Abel:2019zou}. At the classical level, it has been proved that the linearized IDG can yield non-singular static solutions for point sources (including electromagnetic charges), p-branes (and cosmic strings), spinning ring distributions, sources describing NUT charges, null sources, and accelerated particles \cite{Biswas:2011ar,Buoninfante:2018stt,Frolov:2015usa,Boos:2018bxf,Buoninfante:2018xif,Kolar:2020bpo,Boos:2020ccj,Kolar:2021oba}. It was also shown that the linearized IDG prevents mini-black-hole production for small masses \cite{Frolov:2015usa,Frolov:2015bia,Frolov:2015bta}. Also, hints of this non-singular behaviour for astrophysical objects have been studied in \cite{Koshelev:2017bxd,Buoninfante:2019swn}. Exact gravitational waves generated by null sources in full IDG were found in \cite{Kilicarslan:2019njc,Dengiz:2020xbu,Kolar:2021rfl,Kolar:2021uiu}.

Several cosmological studies were performed within IDG, in particular, the possibility of describing inflation \cite{Koshelev:2017tvv,SravanKumar:2018dlo,Koshelev:2020foq,Koshelev:2020xby} and the existence of non-singular bouncing solutions \cite{Biswas:2005qr,Biswas:2010zk,Koshelev:2012qn,Biswas:2012bp,Kumar:2020xsl,SravanKumar:2021xhc,Kumar2021,frolov2021bouncing}. All the previous examples made an extensive use of the recursive ansatz on the Ricci scalar $R$, i.e. ${\Box R= r_1 R+r2}$, which reduce the field equations to local second order non-linear differential equations. As a matter of fact, in \cite{Koshelev:2017tvv} it was proved that such a recursive ansatz for flat FRLW already incorporates all solutions of the IDG field equations with traceless energy-momentum tensor. In this paper, we go beyond the recursive ansatz and find a bouncing solution whose energy-momentum tensor is not traceless and shows much stronger dependence on the non-local form factor. 

The paper is organised as follows: In Section~\ref{sec:evaluationoffeq} we derive the field equations of IDG and explain the treatment of the non-local equations with the spectral analysis of the wave operator. In Section~\ref{sec:nlbouncing} we present a bouncing solution for which the eigenvalue problem can be solved and derive an explicit form of the energy-momentum tensor. In Section~\ref{sec:pf} we find sufficient conditions for the form factor that ensure that the energy density of the perfect fluid is non-negative and SEC is met. We also show that this bouncing behaviour is not possible neither in GR nor in `$R+R^2$ gravity' sourced by a fluid meeting such conditions. Moreover, we find that the non-locality only affects the bounce region, while recovering GR at late times. Finally, we provide a brief summary of the results in Section \ref{sec:conclusions}.

\section{Evaluation of non-local field equations}
\label{sec:evaluationoffeq}

In this paper we consider a non-local theory described by the action of IDG
\begin{equation}\label{eq:action}
\mathsf{S}[\bs{g}]=\frac12\int_M \mathfrak{g}^{\frac12}\left(\varkappa^{-1}(R-2\Lambda)+RF\left(\Box\right)R\right)+\mathsf{S}_{\textrm{m}}\;,
\end{equation}
where $R$ is the Ricci scalar, $\mathfrak{g}^{\frac12}=\sqrt{-g}\,dx^{4}$ is the volume element, ${\varkappa=8\pi G_{\mathrm{N}}}$ is the Einstein gravitational constant, $\Box$~is the wave operator, $\Lambda$ is the cosmological constant, and $\mathsf{S}_{\textrm{m}}$ represents the matter sector of the action. The non-local operator $F\left(\Box\right)$ is ofter referred to as the \textit{form-factor}. It is given by an analytic non-polynomial function\footnote{Here the $f_n$ are dimensionfull constant coefficients}
\begin{equation}
F\left(\Box\right)=\sum_{n=0}^\infty f_{n}\Box^{n}\;.
\end{equation}

Performing the variation with respect to the metric $\bs{g}$, we find the field equations of IDG \cite{Biswas:2005qr,Biswas:2013cha},
\begin{equation}\label{eq:fieldeq}
    \varkappa^{-1}\left(\bs{Ric}-\tfrac12 R\bs{g}+\Lambda\bs{g}\right)+\bs{\Upsilon}+\tfrac12 \bs{g}\big(\Omega+{\Theta}\big)-\bs{\Omega}=\bs{T}\;,
\end{equation}
where the rank-2 tensors $\bs{\Upsilon}$, $\bs{\Omega}$, and the scalar ${\Theta}$ are given by
\begin{equation}\label{eq:omega}
\begin{aligned}
    \bs{\Upsilon} &=\left[2\bs{Ric}-2\bs{\nabla\nabla}-\tfrac12 \bs{g}R+2\bs{g}\Box\right]F\left(\Box\right)R\;,
    \\
    \bs{\Omega} &\equiv \sum_{n=1}^{\infty}f_{n}\sum_{k=0}^{n-1}\bs{\nabla}\Box^k R\bs{\nabla}\Box^{n-k-1}R\equiv R\overrightarrow{\bs{\nabla}}G(\overleftarrow{\Box},\overrightarrow{\Box})\overleftarrow{\bs{\nabla}}R \;,
    \\
    \Theta & \equiv\sum_{n=1}^{\infty}f_{n}\sum_{k=0}^{n-1}\Box^k R\Box^{n-k} R\equiv  R\tfrac{\overleftarrow{\Box}+\overrightarrow{\Box}}{2}G(\overleftarrow{\Box},\overrightarrow{\Box}) R\;.
\end{aligned}
\end{equation}
The arrows over the covariant derivatives denote left/right actions of the derivative. The operator $G(\overleftarrow{\Box},\overrightarrow{\Box})$ is a non-local bilinear operator which is defined as
\begin{equation}
    G(\overleftarrow{\Box},\overrightarrow{\Box})\equiv\frac{F(\overleftarrow{\Box})-F(\overrightarrow{\Box})}{\overleftarrow{\Box}-\overrightarrow{\Box}}\;.
\end{equation}

The presented field equations are very complicated due to their non-local and non-linear structure. There exist three main approaches to solving these equations in the literature:
\begin{itemize}
    \item \underline{Linearized regime \cite{Biswas:2011ar}:} Performing the first order perturbations with respect to a given background, one can approximate the field equations by \textit{linear} non-local equations.
    \item \underline{TN/TIII ansatz \cite{Kolar:2021rfl}:} In certain class of geometries, such as the almost universal spacetimes TN/TIII, the non-linear terms may not contribute even without making any approximations. The field equation may reduce to the \textit{linear} non-local differential equations.
    \item \underline{Recursive ansatz \cite{Biswas:2005qr}:} Assuming the recursive ansatz for the curvature, such as $\Box R=r_1 R+r_2$, the field equations can be recast to the form of \textit{local} (second order) non-linear differential equations.
\end{itemize}
Only the second and third methods produce exact solutions of the full theory, however, such geometries are very special. In the TN/TIII ansatz, the non-linear terms in the curvature do not contribute. In the recursive ansatz, the dependence on the form-factor is very weak in the sense that only the values $F(r_1)$ and $F'(r_1)$ are actually important for the energy-momentum tensor. In this paper, we study a geometry that does not fall in any of these categories. Its energy-momentum tensor has a strong dependence on the non-local form-factor and the non-linear terms do not automatically vanish. 

When solving the linear non-local differential equations, for example those arising in the linearized regime or from the TN/TIII ansatz ansatz, one often employs the spectral representation of the non-local operators using the eigenvalues of the wave operator $\Box$. The linearity then allows one to invert the non-local operators and write the solutions using the associated integral transforms (if the spectrum is continuous). Of course, such an inversion is not possible if the equations are non-linear, however, one can still use the method of spectral decomposition to evaluate the field equation. Namely, we can try to find a certain geometry for which the eigenvalue problem,
\begin{equation}
    \Box \psi_{\lambda} = \lambda \psi_{\lambda}\;,
\end{equation}
possesses simple (analytic) solutions ${\psi_{\lambda}=\psi_{\lambda}(\mathrm{x})}$ for discrete or continuous spectrum $\lambda$. If the eigenfunctions $\psi_{\lambda}$ form a complete orthonormal set, $\langle\psi_{\lambda},\psi_{\lambda'}\rangle=\delta_{\lambda\lambda'}$, then we can decompose a function $f$ in terms of the eigenfunctions~$\psi_{\lambda}$,
\begin{equation}
    f(\mathrm{x})=\sumint_{\lambda} f_{\lambda}\psi_{\lambda}(\mathrm{x})\;,
    \quad
    f_{\lambda}=\langle \psi_{\lambda},f\rangle\;,
\end{equation}
where $\sumint_{\lambda}$ denotes summation $\sum_{\lambda}$ if the spectrum is discrete and integration $\int_{\lambda}$ if the spectrum is continuous. In the latter case, these expressions define direct and inverse integral transforms. (The simplest example is the Fourier transform in flat spacetime in Lorentzian coordinates.)
Equipped with this, we can represent the action of the non-local operator $F(\Box)$ by
\begin{equation}
    F(\Box)f=\sumint_{\lambda} f_{\lambda}F(\Box)\psi_{\lambda}=\sumint_{\lambda} f_{\lambda}F(\lambda)\psi_{\lambda}\;.
\end{equation}
In the next section we present a spacetime, where this process of evaluation of the non-local field equations is tractable, yet still non-trivial in the sense that the energy-momentum tensor depends on all values of $F$.

\section{New non-local bouncing geometries}
\label{sec:nlbouncing}
The search for a geometry for which the field equations can be evaluated using the method described above is quite non-trivial for several reasons: i) There are very few physically interesting geometries that actually admit exact closed-form solution of the eigenvalue problem. ii) In many of these cases the eigenfunctions are either too complicated or do not form a complete set. iii) Even if this is all satisfied, the Ricci scalar may still fail to be decomposable. To deal with these issues, we focus on a flat FLRW ansatz 
\begin{equation}\label{eq:metric}
    \bs{g} = -\bs{\dd}t\bs{\dd}t+a^2\bs{q}\;,
\end{equation}
where ${a=a(t)}$ denotes the \textit{scale factor}, and $\bs{q}$ is the metric of the 3-dimensional flat space. When acting on scalar functions, the wave operator $\Box$ reads
\begin{equation}
    \Box=-\pp_t^2-3H\pp_t\;,
\end{equation}
where ${H=\dot{a}/a}$ is the Hubble parameter. The Ricci tensor $\bs{Ric}$ and the Ricci scalar $R$ of the metric \eqref{eq:metric} are given by the expressions
\begin{equation}
\begin{aligned}
    \bs{Ric} &=-\frac{3\ddot{a}}{a}\bs{\dd}t\bs{\dd}t+(a\ddot{a}+2\dot{a}^2)\bs{q}
    \\
    &=-3(\dot{H}+H^2)\bs{\dd}t\bs{\dd}t+(\dot{H}+3H^2)a^2\bs{q}\;,
    \\
    R &=6\left(\frac{\ddot{a}}{a}+\frac{\dot{a}^2}{a^2}\right)=6(\dot{H}+2H^2)\;.
\end{aligned}    
\end{equation}
The individual terms in the field equations \eqref{eq:fieldeq} can be expressed as
\begin{equation}
\begin{aligned}
    \varkappa^{-1}\left(\bs{Ric}-\tfrac12 R\bs{g}+\Lambda \bs{g}\right) &=\varkappa^{-1}(3H^2-\Lambda)\bs{\dd}t\bs{\dd}t-\varkappa^{-1}(2\dot{H}+3H^2-\Lambda)a^2\bs{q}\;,
    \\
    \bs{\Upsilon} &=\big[\big(6 H \pp_t- 3\dot{H}\big)F(\Box)R\big]\bs{\dd}t\bs{\dd}t+\big[{-}\big(2 \pp_t^2+ 4 H \pp_t+\dot{H}\big)F(\Box)R\big]a^2\bs{q}\;,
    \\
    \tfrac12 \bs{g}\big(\Omega+{\Theta}\big)-\bs{\Omega} &=-\tfrac{1}{2} (\Theta +\Omega_{ tt})\bs{\dd}t\bs{\dd}t+\tfrac{1}{2} (\Theta -\Omega_{ tt})a^2\bs{q}\;.
\end{aligned}    
\end{equation}
Thus, we can find the formula for the energy-momentum tensor of the flat FLRW spacetime,
\begin{equation}\label{eq:energymomentum}
\begin{aligned}
    \bs{T} &=\big[\varkappa^{-1}(3H^2-\Lambda)+\big(6 H \pp_t- 3\dot{H}\big)F(\Box)R-\tfrac{1}{2} (\Theta +\Omega_{ tt})\big]\bs{\dd}t\bs{\dd}t
    \\
    &\feq+\big[{-}\varkappa^{-1}(2\dot{H}+3H^2-\Lambda)- \big(2 \pp_t^2+ 4 H \pp_t+\dot{H}\big)F(\Box)R+\tfrac{1}{2} (\Theta -\Omega_{ tt})\big]a^2\bs{q}\;.
\end{aligned}
\end{equation}
As mentioned above, the spectral decomposition plays a central role in the evaluation of \eqref{eq:energymomentum}. The differential equation for the eigenvalue problem,
\begin{equation}
    \Box\psi_{\omega}=\lambda_{\omega}\psi_{\omega}\;,
\end{equation}
is remarkably simple in the FLRW geometry \eqref{eq:metric}, especially if we focus on spatially homogeneous eigenfunctions ${\psi_{\omega}=\psi_{\omega}(t)}$,
\begin{equation}
\label{eq:eigenfunction}
    \ddot{\psi}_{\omega}+3H\dot{\psi}_{\omega}+\lambda_{\omega}\psi_{\omega}=0\;,
\end{equation}
where the dot stands for the derivative with respect to $t$. Here, $\lambda_{\omega}$ denotes the eigenvalues, labelled by a parameter ${\omega}$. 

In order to find explicit solutions of this equation, one has to specify the geometry even further. Let us focus on the FLRW spacetimes with the following scale factor
\begin{equation}\label{eq:scalefact}
    a=a_0\cosh^{\gamma}(\alpha t)\;,
\end{equation}
with ${\alpha>0}$ and ${\gamma>0}$. This scale factor describes bouncing cosmologies with the bounce at ${t=0}$, which approach de Sitter spacetime with the Ricci scalar ${R=12\alpha^2\gamma^2}$ for ${t\to\pm\infty}$. Note that by explicitly choosing a non-singular geometries we avoid problems with introducing non-local operators that can act on functions living on manifolds with boundaries. The Hubble parameter $H$ for this scale factor is given by
\begin{equation}\label{eq:hubble}
    H={\gamma} \alpha  \tanh (\alpha t)\;.
\end{equation}
The special value ${{\gamma}=1}$, has been studied in the literature \cite{Biswas:2005qr}. An important property of such a spacetime is that the Ricci scalar satisfy recursive ansatz. In other words, the Ricci scalar shifted by a constant is itself an eigenfunction of $\Box$ (i.e., satisfying \eqref{eq:eigenfunction}) because
\begin{equation}
    \Box R = 2\alpha^2 R -24\alpha^4
    \;\Longleftrightarrow\;
    \Box \mathring{R}=2 \alpha^2\mathring{R}\;, 
    \quad 
    \mathring{R}\equiv R-12\alpha^2\;,
\end{equation}
meaning that ${\psi_0=\mathring{R}}$ and ${\lambda_0=2 \alpha^2}$. As mentioned above, the recursive ansatz has proved useful in generating solutions of non-local gravity because it effectively reduces the field equations to the local equations with only a weak dependence on the non-local form-factor.

We will study the case ${{\gamma}=2/3}$, which goes beyond the recursive ansatz, since
\begin{equation}
    \Box R =-\frac{3}{2} \left(R-\frac{16}{3}\alpha^2\right)^2\;.
\end{equation}
The choice ${{\gamma}=2/3}$ is rather unique, since the eigenfunctions take a surprisingly simple form,
\begin{equation}\label{eq:eigenfunc}
    \psi_{\omega}=\frac{1}{\sqrt{2\pi}}\sech(\alpha  t)e^{-i\omega t}\;,
\end{equation}
with the corresponding eigenvalues ${\lambda_{\omega}=\omega^2{+}\alpha^2}$; therefore
\begin{equation}
    \Box \psi_{\omega}=(\omega^2{+}\alpha^2)\psi_{\omega}\;.
\end{equation}
The eigenfunctions \eqref{eq:eigenfunc} are very convenient because they depend on $\omega$ only through the multiplicative factor $e^{-i\omega t}$. Thus, they give rise to the integral transform $\mathsf{G}$ that is closely related to the Fourier transform $\mathsf{F}$,
\begin{equation}
    \mathsf{F}[h](\omega)=\frac{1}{\sqrt{2\pi}}\int_{\mathbb{R}}\!\!d t\,h(t)e^{i\omega t}\;,
    \quad
    \mathsf{F}^{-1}[h](t)=\frac{1}{\sqrt{2\pi}}\int_{\mathbb{R}}\!\!d\omega\,h(\omega)e^{-i\omega t}\;.
\end{equation}
In particular, for the inverse transform of $\mathsf{G}$ we find
\begin{equation}
    f(t)=\mathsf{G}^{-1}[f_{\omega}](t)=\int_{\mathbb{R}}\!\!d\omega\,f_{\omega}\psi_{\omega}(t)=\sech{(\alpha t)}\frac{1}{\sqrt{2\pi}}\int_{\mathbb{R}}\!\!d\omega\,f_{\omega}e^{-i\omega t}=\sech{(\alpha t)}\mathsf{F}^{-1}[f_{\omega}](t)\;.
\end{equation}
Thanks to this relation, we can find also the direct integral transform,
\begin{equation}
    f_{\omega}=\mathsf{G}[f(t)](\omega)=\mathsf{F}[\cosh(\alpha t)f(t)](\omega)=\frac{1}{\sqrt{2\pi}}\int_{\mathbb{R}}\!\!dt\,\cosh(\alpha t)f(t)e^{i\omega t}=\int_{\mathbb{R}}\!\!dt\,\cosh^2(\alpha t)\psi^*_{\omega}(t)f(t)=\langle\psi_{\omega},f\rangle\;,
\end{equation}
and identify the inner product in which $\psi_{\omega}$ are orthonormal. Hence, we are ensured that the integral transform $\mathsf{G}$ that is given by eigenfunctions $\psi_{\omega}$ is well defined and has a corresponding inverse $\mathsf{G}^{-1}$. 

To evaluate the non-local terms \eqref{eq:omega} in the field equations, we have to express the Ricci scalar $R$ with the help of the integral transform $\mathsf{G}$. The Ricci scalar of the flat FLRW with the scale factor \eqref{eq:scalefact} and ${{\gamma}=2/3}$ is
\begin{equation}\label{eq:ricciscalarbounc}
    R=\frac{4}{3}\alpha^2\left(4-\sech^2{\alpha t}\right)\;.
\end{equation}
Apparently, a problem arises here, since the relevant integral diverges due to the constant part of the expression. Instead, we can decompose the Ricci scalar shifted by this constant,
\begin{equation}\label{eq:Riccitransform}
    \hat{R}\equiv R-\frac{16}{3}\alpha^{2}=\mathsf{G}^{-1}[\hat{R}_{\omega}](t)=\int_{\mathbb{R}}d\omega\hat{R}_{\omega}\psi_{\omega}\;,
\end{equation}
where
\begin{equation}
    \hat{R}_{\omega} =\mathsf{G}[\hat{R}(t)](\omega)=\mathsf{F}[\cosh(\alpha t)\hat{R}(t)](\omega)=-\frac{2\sqrt{2\pi}}{3}\alpha\sech\left(\frac{\pi\omega}{2\alpha}\right)\;.
\end{equation}
As we shall see, this decomposition is sufficient to rewrite the field equations to the integral form. Indeed, if we apply the integral representation of the Ricci scalar \eqref{eq:Riccitransform}, we can write
\begin{equation}
\begin{aligned}
    F(\Box)R &=\tfrac{16}{3}\alpha^{2}f_0+\int_{\mathbb{R}} \!\! d\omega \, F(\omega^2{+}\alpha^2)\hat{R}_{\omega}\psi_{\omega}\;,
    \\
    \Omega_{tt} &= \iint_{\mathbb{R}^2} \!\! d\omega d\omega' \,  G(\omega^2{+}\alpha^2,\omega'^2{+}\alpha^2)\hat{R}_{\omega}\hat{R}_{\omega'}\dot{\psi}_{\omega}\dot{\psi}_{\omega'}\;,
    \\
    \Theta &=-\tfrac{16}{3}\alpha^{2}f_{0}R+\left(\tfrac{16}{3}\alpha^{2}\right)^{2}f_{0}+\tfrac{16}{3}\alpha^{2}\int_{\mathbb{R}}\!\!d\omega\,F(\omega^{2}{+}\alpha^{2})\hat{R}_{\omega}\psi_{\omega}
    \\
    &\feq+\iint_{\mathbb{R}^2} \!\! d\omega d\omega' \, \tfrac{\omega^2+\omega'^2+2\alpha^2}{2}G(\omega^2{+}\alpha^2,\omega'^2{+}\alpha^2)\hat{R}_{\omega}\hat{R}_{\omega'} \psi_{\omega}\psi_{\omega'}\;.
\end{aligned}    
\end{equation}
Consequently, the individual terms in \eqref{eq:energymomentum} can be expressed as follows:
\begin{equation}
\begin{aligned}
    \varkappa^{-1}(3H^2-\Lambda) &=\varkappa^{-1} \left(\tfrac{4}{3} \alpha ^2 \tanh ^2(\alpha  t)-\Lambda\right)\;,
    \\
    {-}\varkappa^{-1}(2\dot{H}+3H^2-\Lambda) &=\varkappa^{-1}\left(-\tfrac{4 }{3 }\alpha ^2+\Lambda\right)\;,
    \\
    \big(6H\partial_{t}-3\dot{H}\big)F(\Box)R&=-16\alpha^{2}f_{0}\dot{H}+\int_{\mathbb{R}}\!\!d\omega\,\left[-4i\alpha\tanh(\alpha t)\,\omega-2\alpha^{2}\left(1+\tanh^{2}(\alpha t)\right)\right]F(\omega^{2}{+}\alpha^{2})\hat{R}_{\omega}\psi_{\omega}\;,
    \\
    -\big(2\pp_{t}^{2}+4H\pp_{t}+\dot{H}\big)F(\Box)R&=-\tfrac{16}{3}\alpha^{2}f_{0}\dot{H}+\int_{\mathbb{R}}\!\!d\omega\,\left[2\omega^{2}-\tfrac{4}{3}i\alpha\tanh(\alpha t)\,\omega+\tfrac{2}{3}\alpha^{2}\left(1+\sech^{2}(\alpha t)\right)\right]F(\omega^{2}{+}\alpha^{2})\hat{R}_{\omega}\psi_{\omega}\;,
    \\
    \mp\tfrac12(\Theta\pm\Omega_{tt}) &=\mp\frac12\left[\left(\tfrac{16}{3}\alpha^{2}\right)^{2}f_{0}-\tfrac{16}{3}\alpha^{2}f_{0}R+\tfrac{16}{3}\alpha^{2}\int_{\mathbb{R}}\!\!d\omega\,F(\omega^{2}{+}\alpha^{2})\hat{R}_{\omega}\psi_{\omega}\right]
    \\
    &\feq\mp\iint_{\mathbb{R}^2} \!\! d\omega d\omega' \, \left[\tfrac14(\omega\mp\omega')^2\pm \tfrac{1}{2}i\alpha\tanh(\alpha t)(\omega+\omega')+\tfrac{1}{2}\alpha^2\left(1\pm\tanh^2(\alpha t)\right)\right]
    \\
    &\feq\times G(\omega^2{+}\alpha^2,\omega'^2{+}\alpha^2)\hat{R}_{\omega}\hat{R}_{\omega'} \psi_{\omega}\psi_{\omega'}\;.
\end{aligned}
\end{equation}
When put together, we finally arrive at the explicit formula for the energy-momentum tensor,
\begin{equation}\label{eq:energymomentumfinal}
\begin{aligned}
    \bs{T} &=\bigg[\varkappa^{-1}\left(\tfrac{4}{3}\alpha^{2}\tanh^{2}(\alpha t)-\Lambda\right)-\tfrac{128}{9}\alpha^{4}f_{0}\sech^{2}\left(\alpha t\right)
    \\
    &\feq+\int_{\mathbb{R}}\!\!d\omega\,\left[-4i\alpha\tanh(\alpha t)\,\omega-2\alpha^{2}\left(\tfrac{7}{3}+\tanh^{2}(\alpha t)\right)\right]F(\omega^{2}{+}\alpha^{2})\hat{R}_{\omega}\psi_{\omega}
    \\
    &\feq-\iint_{\mathbb{R}^{2}}\!\!d\omega d\omega'\,\left[\tfrac{1}{4}(\omega-\omega')^{2}+\tfrac{1}{2}i\alpha\tanh(\alpha t)(\omega+\omega')+\tfrac{1}{2}\alpha^{2}\left(1+\tanh^{2}(\alpha t)\right)\right] G(\omega^{2}{+}\alpha^{2},\omega'^{2}{+}\alpha^{2})\hat{R}_{\omega}\hat{R}_{\omega'}\psi_{\omega}\psi_{\omega'}\bigg]\bs{\dd}t\bs{\dd}t
    \\
    &\feq+\bigg[\varkappa^{-1}\left(-\tfrac{4}{3}\alpha^{2}+\Lambda\right)+\int_{\mathbb{R}}\!\!d\omega\,\left[2\omega^{2}-\tfrac{4}{3}i\alpha\tanh(\alpha t)\,\omega+\tfrac{2}{3}\alpha^{2}\left(5+\sech^{2}(\alpha t)\right)\right]F(\omega^{2}{+}\alpha^{2})\hat{R}_{\omega}\psi_{\omega}
    \\
    &\feq+\iint_{\mathbb{R}^{2}}\!\!d\omega d\omega'\,\left[\tfrac{1}{4}(\omega+\omega')^{2}-\tfrac{1}{2}i\alpha\tanh(\alpha t)(\omega+\omega')+\tfrac{1}{2}\alpha^{2}\left(1-\tanh^{2}(\alpha t)\right)\right] G(\omega^{2}{+}\alpha^{2},\omega'^{2}{+}\alpha^{2})\hat{R}_{\omega}\hat{R}_{\omega'}\psi_{\omega}\psi_{\omega'}\bigg]a^2\bs{q}\;.
\end{aligned}
\end{equation}
In the next section we will study its properties and show that the specific non-singular geometry we chose, i.e. \eqref{eq:scalefact} with $\gamma=2/3$, can be actually sourced by a perfect fluid with non-negative energy density that meets SEC.

\section{Properties of the perfect fluid}
\label{sec:pf}

The energy-momentum tensor \eqref{eq:energymomentumfinal} strongly depends on the form-factor $F$ through expressions involving $F(\omega^2+\alpha^2)$ and $G(\omega^2+\alpha^2,\omega'^2+\alpha^2)$ integrated over all values of $\omega$ and $\omega'$. In other words, we have a greater freedom to choose a form-factor $F$ such that the energy-momentum tensor for a given geometry meets the desired properties. We shall first require that the energy density is non-negative in order for the perfect fluid to be physically reasonable. Secondly, we shall impose that the perfect fluid satisfies SEC. Note that this implies that a geometry that would be sourced by the same perfect fluid in GR would be singular.

This section is divided in two parts: In the first part, we prove that neither in GR nor in `${R+R^2}$ gravity' it is possible to meet the requirements on the energy-momentum tensor of the perfect fluid for this geometry. In the second part, we study these requirements in the non-local theory. In particular, we find some sufficient conditions on the form-factors and also provide a specific numerical example of a form-factor that fulfills these conditions.

\subsection{Local gravity: GR \& \texorpdfstring{$R+R^2$}{R+R2}}
The energy-momentum tensor of the perfect fluid in the flat FLRW spacetime is given by
\begin{equation}
    \bs{T} =\rho\,\bs{\dd}t\bs{\dd}t +p\, a^2\,\bs{q}\;,
\end{equation}
where $\rho$ is the energy density and $p$ is the pressure. Let us see whether the bouncing cosmology \eqref{eq:scalefact} with ${{\gamma}=2/3}$ can be sourced by a physical matter. First we consider the general relativity, i.e,  ${F(\Box)=0}$. The condition on non-negative energy density reads
\begin{equation}\label{eq:GRFE}
    \rho=\varkappa^{-1}\left(\tfrac{4}{3}\alpha^{2}\tanh^{2}(\alpha t)-\Lambda\right)\geq 0\;,
\end{equation}
which is satisfied for all $t$ only if ${\Lambda\leq 0}$. Next, we demand fulfilment of SEC,
\begin{equation}\label{eq:SECgr}
\begin{aligned}
    \rho+p &=-\tfrac{4}{3}\varkappa^{-1}\alpha^2\sech^{2}(\alpha t)\geq0\;,
    \\
    \rho+3p &=\varkappa^{-1}\left(\tfrac{4}{3}\alpha^{2}\tanh^{2}(\alpha t)-4\alpha^{2}+2\Lambda\right)\ge 0\;.
\end{aligned}
\end{equation}
As we can see, the first condition can never hold and the second one is met for all $t$ only if ${\Lambda\geq 2\alpha^2}$. Therefore, the perfect fluid generating this geometry in GR does not meet SEC. (A special example is pure GR without the cosmological term, ${\Lambda=0}$, where the violation of SEC can be also argued based on the Hawking theorem applied to this non-singular bouncing FLRW spacetime.) Before we proceed to another theory, let us investigate what happens at late/early times ${t\to\pm\infty}$ (i.e., far from the bounce ${t=0}$) . The energy density \eqref{eq:GRFE} and SEC \eqref{eq:SECgr} turn into ${\Lambda\leq 4\alpha^{2}/3}$ and ${\Lambda\geq 4\alpha^{2}/3}$, respectively. This means that both conditions are asymptotically satisfied only if ${\Lambda= 4\alpha^{2}/3}$, which corresponds to the asymptotic vacuum, ${\bs{T}\to 0}$ for ${t\to\pm\infty}$. For this specific choice ${\Lambda=4\alpha^{2}/3}$, the cosmological term accounts for the accelerated expansion, where the vacuum corresponds to de Sitter spacetime with the Ricci scalar ${R=16\alpha^{2}/3}$.

Let us investigate if this behavior can be improved in `${R+R^2}$ gravity', which is given by a constant form-factor ${F(\Box)=f_0}$. The energy density is non-negative if
\begin{equation}\label{eq:FER2}
    \rho=\varkappa^{-1}\left(\tfrac{4}{3}\alpha^{2}\tanh^{2}(\alpha t)-\Lambda\right)-8f_0\alpha^{4}\sech^{4}(\alpha t)\geq 0\;,
\end{equation}
which holds for all $t$ if ${\Lambda\leq-8\varkappa f_0 \alpha^4}$. SEC reads
\begin{equation}\label{eq:SEC2}
\begin{aligned}
    \rho+p &=-\tfrac{4}{3}\varkappa^{-1}\alpha^2\sech^{2}(\alpha t)-16 f_0\alpha^4\sech^4(\alpha t)\ge 0\;,
    \\
    \rho+3p &=\varkappa^{-1}\left(\tfrac{4}{3}\alpha^{2}\tanh^{2}(\alpha t)-4\alpha^{2}+2\Lambda\right)-32 f_0\alpha^4\sech^4(\alpha t)\ge 0\;.
\end{aligned}
\end{equation}
These two conditions are met for all $t$ only if $f_0\leq-\varkappa^{-1}/12\alpha^2$, and ${\Lambda\geq 2\alpha^2+ 16\varkappa f_0 \alpha^4}$, respectively. Unfortunately, $f_0$ should be positive to avoid Dolgov--Kawasaki instability \cite{Dolgov:2003px,Faraoni:2006sy}, which obviously violates SEC. Taking the limits ${t\to\pm\infty}$ of \eqref{eq:FER2} and \eqref{eq:SEC2} we again arrive at ${\Lambda\leq 4\alpha^{2}/3}$ and ${\Lambda\geq 4\alpha^{2}/3}$, respectively. As in GR, both conditions are asymptotically satisfied only for ${\Lambda= 4\alpha^{2}/3}$ corresponding to the asymptotic vacuum. This is because the contributions to the energy-momentum tensor arising from $R^2$ term in the action decay faster than the contributions from the Einstein--Hilbert term.

\subsection{Non-local gravity: IDG}

Let us see if the non-local theory described by the action of IDG \eqref{eq:action} can improve the properties of the perfect fluid and give rise to a physically reasonable matter. Our aim is to find sufficient conditions for the form-factor under which the perfect fluid meets the required properties. First, recall that the bouncing geometry tends to de Sitter spacetime for ${t\to\pm\infty}$. Its Hubble parameter and Ricci scalar approach constants ${H=2\alpha/3}$ and ${R=16\alpha^{2}/3}$ with exponential rate, see \eqref{eq:hubble} (with $\gamma=2/3$) and \eqref{eq:ricciscalarbounc}. Moreover, the derivatives ${\pp_t^k H}$ and ${\pp_t^k R}$ for ${k>0}$ decay exponentially. Inspecting \eqref{eq:energymomentum}, we see that the contributions to the energy-momentum tensor from all terms except for the Einstein--Hilbert term become negligible for ${t\to\pm\infty}$, and thus IDG effectively reduces to GR. This has an important implication for the requirements on the perfect fluid because, as we have seen, they can be satisfied asymptotically only if we set ${\Lambda=4\alpha^{2}/3}$; all other choices of $\Lambda$ would necessarily lead to negative energy density or violation of SEC. 

At this point we should analyze possible form-factors that allow for non-negative energy density and meet SEC. The first restriction, however, comes already from the integral transform we used to evaluate the energy-momentum tensor. All (1- and 2- dimensional) integrals in \eqref{eq:energymomentumfinal}  must converge, which already constraints the form-factors that we can choose. We have to demand that the absolute value of integrands are integrable functions. Therefore, a sufficient condition for convergence of the 1-dimensional integrals over $\mathbb{R}$ is that functions
\begin{equation}
    |\omega|^k\sech\left(\frac{\pi\omega}{2\alpha}\right)\left|F\left(\omega^{2}{+}\alpha^{2}\right)\right|\;, \quad k=0,1,2\;, 
\end{equation}
belong to $L_1(\mathbb{R})$. To find the conditions of convergence, we realize that the integrands are analytic functions ($F$ is analytic), so it is sufficient to investigate the convergence of the integrals over the intervals $(-\infty,c_{-})$ and $(c_{+},\infty)$ for some constants $c_{-}$ and $c_{+}$. We will rely on the limit comparison test for improper integral, which says that $\int_{c_{\pm}}^{\pm\infty}\!\! d\omega\,f(\omega)$ converges if $\int_{c_{\pm}}^{\pm\infty}\!\! d\omega\,g(\omega)$ is a known convergent integral and
\begin{equation}
    \lim_{\omega\to\pm\infty} f(\omega)/g(\omega) =L_{\pm}\;, \quad 0\leq L_{\pm}<\infty\;.
\end{equation}
We shall use the slowly decreasing function (with a convergent improper integral), ${g(\omega)={|\omega|^{-1-\varepsilon}}}$, $\varepsilon>0$. Then, it is clear that the most restrictive condition that we can write among the three mentioned conditions (${k=0,1,2}$) is
\begin{equation}\label{eq:convergencesingle}
    \lim_{\omega\to\pm\infty}|\omega|^{3+\varepsilon}e^{\mp\frac{\pi\omega}{2\alpha}}\left|F\left(\omega^{2}{+}\alpha^{2}\right)\right|=L_{\pm}\;,
    \quad
    0\le L_{\pm}<\infty\;,
    \quad
    \varepsilon>0\;.
\end{equation}

Before we examine the convergence of the 2-dimensional integrals, it is important to convince ourselves that the integrands are also analytic, which may not be apparent because of the presence of the function $G$. Indeed, the function $G(x,y)$ is analytic everywhere including the points ${x=y}$, where it can by analytically extended by its limit ${G\left(x,x\right)=F'\left(x\right)}$. The conditions for the convergence of the 2-dimensional integrals over $\mathbb{R}^2$ can be obtained from the 2-dimensional Fourier transform conditions of existence \cite{goodman2008introduction}, namely the absolute integrability of the integrand. Hence, sufficient conditions for convergence are that functions
\begin{equation}
    |\omega|^k|\omega'|^l\sech\left(\frac{\pi\omega}{2\alpha}\right)\sech\big(\frac{\pi\omega'}{2\alpha}\big)\left|G\left(\omega^{2}{+}\alpha^{2},\omega'^{2}{+}\alpha^{2}\right)\right|\;, \quad k+l=0,1,2\;,
\end{equation}
belong to $L_1(\mathbb{R}^2)$. As we did for the 1-dimensional integrals, we can use the limit test to study the convergence \cite{Ghorpade2010}. We can choose the slowly decreasing function with convergent improper integral over $A_R=\left\{(x,y)\in \mathbb{R}^2\;s.t.\;x^2+y^2\ge R\right\}$, which is  ${g(\omega,\omega')=\left(\omega^{2}+\omega'^{2}\right)^{-1-\varepsilon}}$, ${\varepsilon>0}$. Applying the same reasoning as in the 1-dimensional case, the most restrictive conditions coming from the limit convergence test are now
\begin{equation}\label{eq:convergencedouble}
    \lim_{(\omega,\omega')\to\infty}\left(\omega^{2}+\omega'^{2}\right)^{1+\varepsilon}|\omega|^{k}|\omega'|^{l}e^{-\frac{\pi\left|\omega\right|}{2\alpha}}e^{-\frac{\pi\left|\omega\right|'}{2\alpha}}\left|G\left(\omega^{2}{+}\alpha^{2},\omega'^{2}{+}\alpha^{2}\right)\right|=L\;,\quad k+l=2\;,\quad 0\le L<\infty\;,\quad\varepsilon>0\;.
\end{equation}

Having established the convergence of the integrals, we can now move on to the condition for non-negative energy density ${\rho\ge 0}$ and SEC ${\rho+p\ge 0}$, ${\rho+3p\ge 0}$. Since we want to satisfy all these inequalities at the same time, we can easily see that ${\rho+p\ge 0}$ is actually redundant for our needs: Let us assume that ${\rho\ge 0}$. If $p\ge0$ then SEC holds. On the other hand, if ${p<0}$ than $\rho+3p\ge 0$ automatically implies $\rho+p\ge 0$. Therefore, we only need to deal with the quantities $\rho$ and ${\rho+3p}$, which can be read out from \eqref{eq:energymomentumfinal} (where we set ${\Lambda=4\alpha^{2}/3}$),
\begin{equation}\label{eq:rhorhop3p}
\begin{aligned}
    \rho &=-\left(\tfrac{4}{3}\varkappa^{-1}+\tfrac{128}{9}\alpha^{2}f_{0}\right)\alpha^{2}\sech^{2}(\alpha t)+\int_{\mathbb{R}}\!\!d\omega\,\left[-4i\alpha\tanh(\alpha t)\,\omega-2\alpha^{2}\left(\tfrac{7}{3}+\tanh^{2}(\alpha t)\right)\right]F(\omega^{2}{+}\alpha^{2})\hat{R}_{\omega}\psi_{\omega}
    \\
    &\feq-\iint_{\mathbb{R}^{2}}\!\!d\omega d\omega'\,\left[\tfrac{1}{4}(\omega-\omega')^{2}+\tfrac{1}{2}i\alpha\tanh(\alpha t)(\omega+\omega')+\tfrac{1}{2}\alpha^{2}\left(1+\tanh^{2}(\alpha t)\right)\right]
    \\
    &\feq\times G(\omega^{2}{+}\alpha^{2},\omega'^{2}{+}\alpha^{2})\hat{R}_{\omega}\hat{R}_{\omega'}\psi_{\omega}\psi_{\omega'}\;,
    \\
    \rho+3p &= -\left(\tfrac{4}{3}\varkappa^{-1}+\tfrac{128}{9}\alpha^{2}f_{0}\right)\alpha^{2}\sech^{2}(\alpha t)+\int_{\mathbb{R}}\!\!d\omega\,\left(\tfrac{10}{3}\alpha^{2}+6\omega^{2}+4\alpha^{2}\sech^{2}\left(\alpha t\right)-8i\alpha\omega\tanh(\alpha t)\right)F(\omega^{2}{+}\alpha^{2})\hat{R}_{\omega}\psi_{\omega}
    \\
    &\feq+\tfrac{1}{2}\iint_{\mathbb{R}^{2}}\!\!d\omega d\omega'\,\left[2\alpha^{2}+\omega^{2}+\omega'^{2}+4\omega\omega'+4\alpha\tanh(\alpha t)(-\alpha\tanh(\alpha t)-i(\omega+\omega'))\right]
    \\
    &\feq\times G(\omega^{2}{+}\alpha^{2},\omega'^{2}{+}\alpha^{2})\hat{R}_{\omega}\hat{R}_{\omega'}\psi_{\omega}\psi_{\omega'}\;.
\end{aligned}
\end{equation}

Let us take a look at these expression term by term. If we want to have a positive contribution from the first term of $\rho$, we need
\begin{equation}\label{eq:C1}
    f_{0}<-\frac{3\varkappa^{-1}}{32\alpha^{2}}\;.
\end{equation}
To understand the second term (the 1-dimensional integral), we evaluate an auxiliary integral with the constant positive form-factor ${F(\omega^2+\alpha^2)=C>0}$,
\begin{equation}
    C\int_{\mathbb{R}}\!\!d\omega\,\left[-4i\alpha\tanh(\alpha t)\,\omega-2\alpha^{2}\left(\tfrac{7}{3}+\tanh^{2}(\alpha t)\right)\right]\hat{R}_{\omega}\psi_{\omega}=\tfrac{8}{9}C\alpha^{4}(16\cosh^2(\alpha t)-9)\sech^{4}(\alpha t)\;.
\end{equation}
Since this integral is always positive, we might be tempted to assume that ${F(x)}$ is positive and approximately constant for all $x$ in order to make the second term of $\rho$ also positive. Nevertheless, this obviously contradicts the condition \eqref{eq:C1}. To overcome this issue we propose that $F(x)$ should be almost-constant and positive (${\approx C>0}$) only for ${x>\alpha^2}$ but negative for $x\in(0,\zeta^2)$, with ${\alpha>\zeta>0}$, so that $f_0$ meets the requirement \eqref{eq:C1}. Although the third term (the 2-dimensional integral) might be negative, it should not contribute too much because $G(\omega^2{+}\alpha^2,\omega'^2{+}\alpha^2)$ is almost zero for all ${(\omega,\omega')\in\mathbb{R}^2}$.

Moving on to ${\rho+3p}$, we realize that its first term is the equal to the first term of $\rho$, so it is automatically positive thanks to \eqref{eq:C1}. The second term of ${\rho+3p}$ (the single integral) can be analyzed again by calculating the approximate integral with ${F(\omega^2{+}\alpha^2)=C>0}$,
\begin{equation}\label{eq:Term2}
\begin{aligned}
    C\int_{\mathbb{R}}\!\!d\omega\,\left(\tfrac{10}{3}\alpha^{2}+6\omega^{2}+4\alpha^{2}\sech^{2}\left(\alpha t\right)-8i\alpha\omega\tanh(\alpha t)\right)\hat{R}_{\omega}\psi_{\omega}=\tfrac{32}{9}C\alpha^{4}(4\sinh^2(\alpha t)-5)\sech^{4}(\alpha t)\;,
\end{aligned}
\end{equation}
which is positive only for ${\left|t\right|>\arcsinh \left(\sqrt{5}/2\right)/\alpha}$. Let us see if we can counter the negativity of this term with the positivity of the first term of ${\rho+3p}$. By summing the first term of ${\rho+3p}$ with the negative contribution of \eqref{eq:Term2} only, we obtain
\begin{equation}
    -\left(\tfrac{4}{3}\varkappa^{-1}+\tfrac{128}{9}\alpha^{2}f_{0}\right)\alpha^{2}\sech^{2}(\alpha t)-\tfrac{160}{9}C\alpha^{4}\sech^{4}(\alpha t)\;.
\end{equation}
Since ${\sech^{2}(x)\ge \sech^{4}(x)}$ for all $x$, this expression is positive if
\begin{equation}\label{eq:Cf0}
    f_{0}<-\frac{3\varkappa^{-1}}{32\alpha^{2}}-\frac{5}{4}C\;.
\end{equation}
The third term of ${\rho+3p}$ (the 2-dimensional integral) is negligible for the same reason as the analogous term in the analysis of~$\rho$.

\begin{figure}
\centering
    \includegraphics[width=0.48\textwidth]{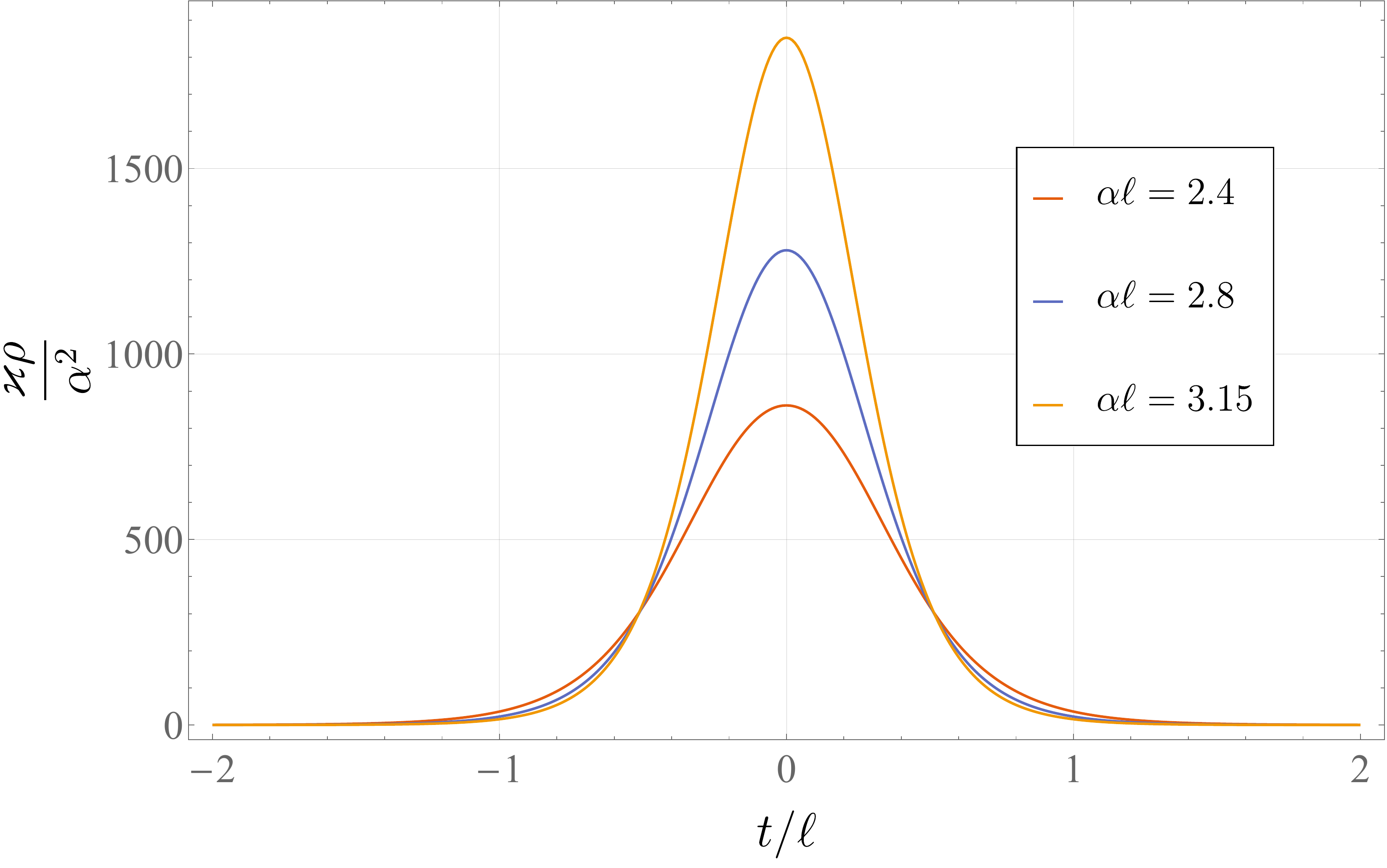}
    \quad
    \includegraphics[width=0.48\linewidth]{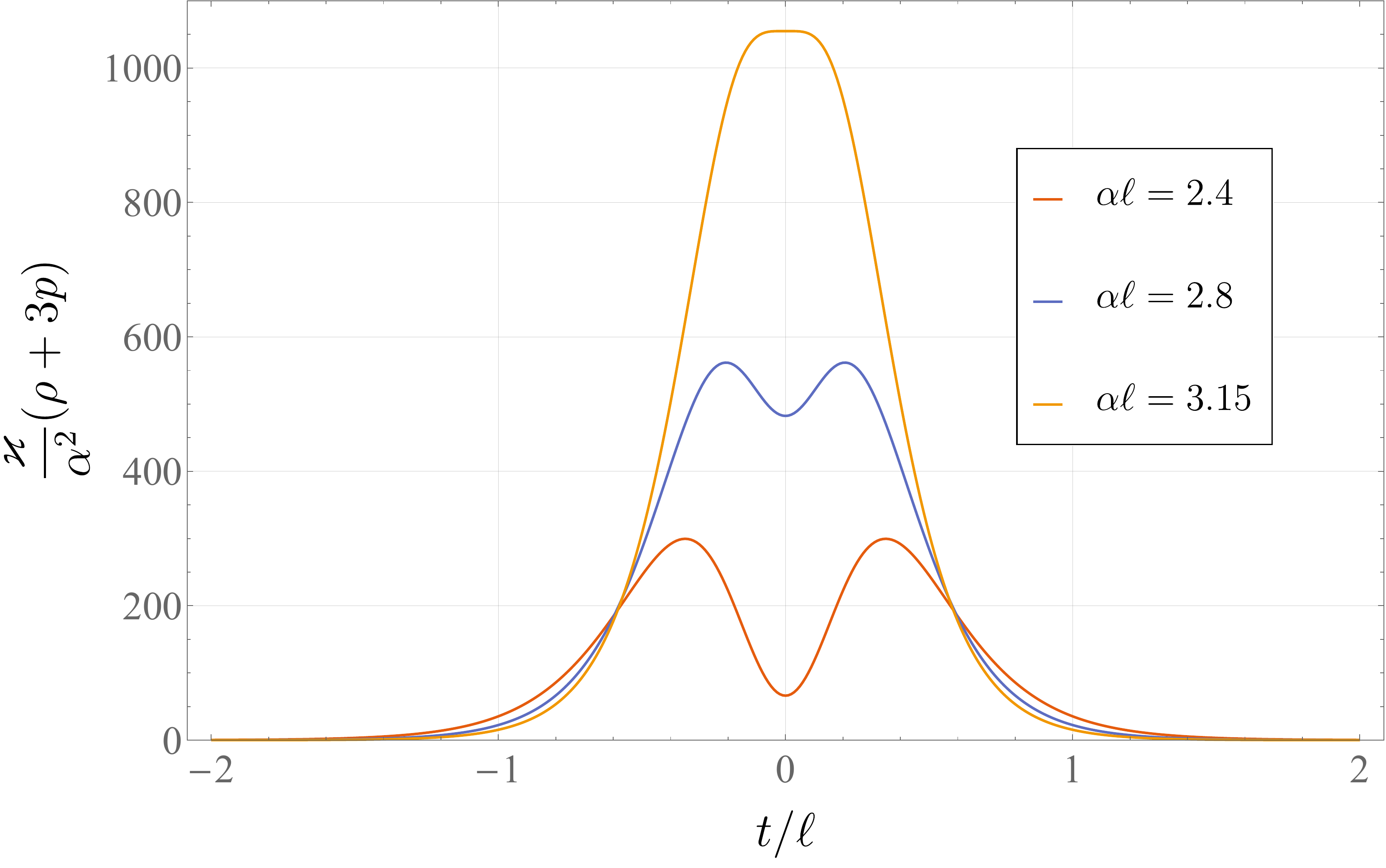}
    \caption{Graphs of the dimensionless quantities ${\varkappa\rho}/{\alpha^2}$ (left) and the quantity ${\varkappa(\rho+3p)}/{\alpha^2}$ (right) of the perfect fluid that sources the bouncing cosmology for the form-factor given in \eqref{eq:FFSEC2}, for three different values of $\alpha\ell$. We have chosen ${C\alpha^2=1.3225}$ and ${\zeta/\alpha=0.3887}$, which satisfy \eqref{eq:cond1} and \eqref{eq:cond2} with ${\epsilon=0.034}$.}
    \label{fig:12}
\end{figure}

To summarize, we have arrived at the following sufficient conditions for the form-factor leading to physically reasonable perfect fluid:
\begin{itemize}
    \item The form-factor must ensure conditions for convergence of integrals appearing in the energy-momentum tensor, i.e., \eqref{eq:convergencesingle} and \eqref{eq:convergencedouble}.
    \item The form-factor must be almost-constant and positive, $F\left(x\right)\approx C>0$, for ${x>\alpha^2}$.
    \item The value of ${f_0\equiv F(0)}$ has to be sufficiently negative to meet condition \eqref{eq:Cf0}\;.
\end{itemize}
In order to verify these finding, we shall provide an explicit example of a form-factor that gives rise to non-negative energy density and the fulfillment of SEC. Let us consider the following form-factor
\begin{equation}\label{eq:FFSEC2}
    F(\Box)=C-Ce^{-\ell^{2}(\Box-\zeta^2)}\;,
\end{equation}
for which we have ${f_{0}=C-C e^{\ell^{2}\zeta^2}}$. This form-factor has been constructed to recover the local limit, i.e., ${F(\Box)=0}$, for ${\ell\to 0}$. First, let us note that the function \eqref{eq:FFSEC2} obeys the convergence conditions \eqref{eq:convergencesingle} and \eqref{eq:convergencedouble} with ${L_{\pm}=0}$ and ${L=0}$, respectively. Next, we notice that it quickly approaches a constant value. To ensure that $F(x)$ is almost-constant for ${x>\alpha^2}$ we propose
\begin{equation}\label{eq:cond1}
    e^{\ell^{2}(\zeta^2-\alpha^{2})}<\epsilon\;,
\end{equation}
where ${\epsilon>0}$ is a sufficiently small number. Finally, we make $f_0$ negative enough according to \eqref{eq:Cf0},
\begin{equation}\label{eq:cond2}
    e^{\ell^{2}\zeta^2}>\frac34\left(\frac{\varkappa^{-1}}{8\alpha^{2}C }+3\right)\;.
\end{equation}
For a given constant ${\alpha}$ we can always find the values of $C$, $\ell$, and $\zeta$ so that the constraints \eqref{eq:cond1} and \eqref{eq:cond2} are satisfied to a prescribed value of $\epsilon$. The graph of the functions ${\rho}$ and ${\rho+3p}$ computed numerically from \eqref{eq:rhorhop3p} are plotted in Figure~\ref{fig:12} in dimensionless quantities. We can see that the energy density is positive and SEC is satisfied everywhere. Hence, in this non-local modification one can obtain a bouncing cosmology sourced by a physically relevant perfect fluid. Finally, it is interesting to note that for this particular form factor \eqref{eq:FFSEC2}, we recover the $R-\beta R^2$ gravity for large $\alpha \ell$, which we have rendered as an unphysical theory due to its instabilities. In order to have a non-local behaviour that fulfills SEC we need to choose $\ell$ to be of the same order as $\alpha$, as can be seen in Figure \ref{fig:12}. This means that the non-locality scale should be of the same order as the de Sitter radius $r_{\textrm{dS}}$, because $r_{\textrm{dS}}=\sqrt{3/\Lambda}=\frac{3}{2}\alpha^{-1}$. Finally, it is interesting to note that the effects of non-locality are only appreciable around the bounce, and that GR is recovered as one distances from it. 

\section{Conclusions}\label{sec:conclusions}

In this paper, we found a new cosmological bouncing geometry for which the field equations can be evaluated using the spectral decomposition with respect to the eigenfunctions of the wave operator. Furthermore, dependence of the resulting energy-momentum tensor \eqref{eq:energymomentumfinal} on the form-factor is much stronger than for the previous solutions available in the literature. Since the new solution does not satisfy the recursive ansatz, ${\Box R = r_1 R +r_2}$, the Ricci scalar (after a constant shift) is no longer an eigenfunction on its own, but it still admits an eigenfunction decompositions. The geometry we choose is rather special in the sense that the eigenvalue problem of the wave operator in this background is exactly solvable and the corresponding eigenfunctions are rather simple.

We showed that due to stronger dependence of the energy-momentum on the form-factor $F(\Box)$, one can tune it in such a way that: i) the integrals appearing in the formula for the energy-momentum tensor converge ii) the solution is sourced by a perfect fluid with non-negative energy density iii) the perfect fluid satisfies SEC. One of the simplest form-factors satisfying i)-iii) can be constructed rather straightforwardly by taking $F(x)$ to be sufficiently negative at ${x=0}$ and almost constant and positive in the domain of integration. We have also shown that the effects of non-locality are only relevant around the bounce, and that GR is recovered at late times.   

A natural extension of this work is to find other types of form-factors satisfying i)-iii) (especially those for which ${F(0)\ge 0}$). Another interesting but rather involved problem is the perturbative stability of the solution, which may put further restrictions on the form-factor. It will also be worth exploring more general solutions for which the eigenvalue problem of the d'Alembertian operator can be solved.


\section*{Acknowledgements}

I.K. and A.M. were supported by Netherlands Organization for Scientific Research (NWO) grant no. 680-91-119. F.J.M.T. acknowledges financial support from NRF grants no. 120390, reference: BSFP190416431035; no. 120396, reference: CSRP190405427545; no. 101775, reference: SFH150727131568.

\bibliography{references}

\end{document}